\documentclass[aps,prd,twocolumn,nofootinbib,superscriptaddress]{revtex4}

\usepackage{graphicx} % Required for inserting images
\usepackage{physics}
\usepackage{amssymb}
\usepackage{amsmath}
\usepackage{hyperref}
\usepackage[capitalize]{cleveref}
\usepackage{float} % here for H placement parameter

\usepackage{aas_macros}

\usepackage{xcolor}

\newcommand{\Rlisa}{\mathcal{R}_\mathrm{LISA}}
\newcommand{\beam}{\theta_\mathrm{m}}
\newcommand{\flow}{f_\mathrm{low}}
\newcommand{\fhigh}{f_\mathrm{high}}
\newcommand{\bup}{\beta_\mathrm{up}}
\newcommand{\snrcut}{\textup{SNR}_\mathrm{cut}}
\newcommand{\Tobs}{T_\mathrm{obs}}

\graphicspath{
    {figures/}
}

\begin{document}

\title{Cosmic string bursts in LISA}

\author{Pierre Auclair}
\email{pierre.auclair@uclouvain.be}
\affiliation{Cosmology, Universe and Relativity at Louvain (CURL),
Institute of Mathematics and Physics, University of Louvain, 2 Chemin
du Cyclotron, 1348 Louvain-la-Neuve, Belgium}

\author{Stanislav Babak}
\email{stas@apc.in2p3.fr}
\affiliation{Universit\'e Paris Cit\'e, CNRS, Astroparticule et Cosmologie, F-75013 Paris, France}

\author{Hippolyte Quelquejay Leclere}
\email{quelquejay@apc.in2p3.fr}
\affiliation{Universit\'e Paris Cit\'e, CNRS, Astroparticule et Cosmologie, F-75013 Paris, France}

\author{Dani\`ele A.~Steer}\email{steer@apc.in2p3.fr}
\affiliation{Universit\'e Paris Cit\'e, CNRS, Astroparticule et Cosmologie, F-75013 Paris, France}

\date{\today}

\begin{abstract}

Cosmic string cusps are sources of short-lived, linearly polarised gravitational wave bursts
which can be searched for in gravitational wave detectors.
We assess the capability of LISA to detect these bursts using the latest LISA configuration and operational assumptions. For such short bursts, we verify that LISA  can be considered as ``frozen", namely that one can neglect LISA's orbital motion. We consider two models for the network of cosmic string loops, and estimate that LISA should be able to detect 4-30 bursts per year assuming a string tension $G\mu \approx 10^{-10.6} - 10^{-10.1}$ and  detection threshold $\rm{SNR} \ge 20$. Non-detection of these bursts would constrain the string tension to $G\mu\lesssim 10^{-11}$ for both models.

\end{abstract}

\maketitle

\section{Introduction}

The scientific objectives of the LISA mission \cite{SciRD}, whose launch is planned in 2037, are incredibly broad and cover, amongst other things, the astrophysics of stellar binaries, the detailed properties of black holes and tests of General Relativity, galaxy formation and the measurement of cosmological parameters (see \cite{2019BAAS...51c..34L,2019BAAS...51c..67C,2019BAAS...51c.432C,2019BAAS...51c.109C,2019BAAS...51c.123B,2019BAAS...51c..73N,2019BAAS...51c..42B, 2019BAAS...51c..76C,2019arXiv190304592M,2019BAAS...51c..32B,2019BAAS...51c.175B} for recent white papers). Furthermore, LISA may also discover new cosmological sources of gravitational waves (GW), either through their burst-like signal, or from their contribution to the stochastic GW background (SGWB), or possibly both.  In this paper we focus on one such GW source, namely cosmic strings, which are line-like topological defects that
may be formed in symmetry breaking phase transitions in the early universe \cite{Kibble:1976sj,Hindmarsh:1994re,Vilenkin:2000jqa,Vachaspati:2015cma}. The potential of LISA to detect cosmic strings through their contribution to the SGWB was recently studied in depth in \cite{Auclair:2019wcv}.  However, as is well known, see e.g.~\cite{Damour:2000wa,Damour:2001bk}, cosmic string cusps --- points on the string which instantaneously travel at the speed of light --- also source GW bursts.
Whilst these have been searched for with LIGO-Virgo-Kagra \cite{LIGOScientific:2017ikf,LIGOScientific:2021nrg}, at LISA frequencies the existing studies are somewhat dated and limited to the Mock LISA Data Challenge 3 (MLDC 3.4) \cite{Babak:2008aa,Cohen:2010xd,ShapiroKey:2008ckh, MockLISADataChallengeTaskForce:2009wir}, or do not model the response of LISA to a cosmic string burst~\cite{Cui:2019kkd}.
The aim of this paper is to reconsider the cosmic string burst signature taking the latest LISA configuration and operation assumptions with the most up-to-date cosmic string models.  We do
not deal with the detection of these signals assuming that the techniques similar to \cite{ShapiroKey:2008ckh} are efficient.

We consider standard (non-current carrying) cosmic strings parametrised by their dimensionless energy per unit length $G\mu$
related to the energy scale $\eta$ of the phase transition by
\begin{equation}
	G\mu \sim 10^{-6} \qty(\frac{\eta}{10^{16} ~\mathrm{GeV}})^2.
\end{equation}
A network of cosmic strings contains both infinite strings as well as a population of closed loops \cite{Vilenkin:2000jqa}.  Multiple studies have shown that the  network evolves to an attractor self-similar scaling solution in which the energy density in strings is a fixed fraction of the energy density of the universe, and all characteristic length scales of the string network are proportional to cosmic time $t$.  Whereas the scaling infinite string network leaves imprints at CMB scales \cite{Ringeval:2010ca} with current constraints $G\mu < 10^{-7}$ \cite{Ade:2013xla}, the GW signal is predominantly sourced by the loop distribution. As loops oscillate they decay into GWs, and since loops of different sizes are permanently sourced by the infinite string network (from formation until today), the produced GWs cover decades in frequency.  They can therefore be probed for by LIGO-Virgo-Kagra, LISA, and PTA experiments.
In \cite{LIGOScientific:2017ikf,LIGOScientific:2021nrg}, the LIGO-Virgo-Kagra collaboration has searched for both their SGWB and burst signatures.  The resulting constraints \cite{LIGOScientific:2021nrg} depend on the loop distribution, and are
\begin{align}
 G\mu &\lesssim 9.6 \times 10^{-9} \quad {\text{BOS Model }}\nonumber
 \\
    G\mu &\lesssim 4 \times 10^{-15} \quad \; {\text{LRS Model}}
    \nonumber
\end{align}
where the LRS and BOS Models (the letters correspond to the author's names) refer to the two main loop distributions in the current literature, given in Refs.~\cite{Blanco-Pillado:2013qja} and \cite{Lorenz:2010sm} respectively.
From the SGWB only, at PTA frequencies, the current constraints are $G\mu \lesssim 10^{-10}$~\cite{Blanco-Pillado:2017rnf,Ringeval:2017eww,Ellis:2020ena,Blasi:2020mfx,Bian:2022tju,Chen:2022azo}.
In the LISA frequency band, the SGWB from cosmic strings was recently studied in \cite{Auclair:2019wcv}, where it was shown that LISA should detect the SGWB from strings with $G\mu \gtrsim {\cal {O}}(10^{-17})$.  As stated above, our aim in this paper is to focus on the {\it burst} signature at LISA frequencies.

In section \ref{sec:2} we recall the main properties of the beamed burst signal from cusps, including the frequency dependence of the opening angle of the beam (which is broader at LISA rather than LIGO frequencies, meaning it is a priori easier to detect).
Then in section \ref{sec:snr} the salient features of the LISA response are summarised. We determine the cosmic string burst efficiency, namely the probability that LISA can detect a burst of a given amplitude, \emph{i.e.~}the probability that its SNR is above a given value $\snrcut$.
In Section \ref{sec:rate}, we derive the rate of bursts observable by LISA. We then evaluate the expected rate for the LRS and BOS models in section \ref{sec:5}. We also consider the case in which LISA does not detect bursts from strings during the mission duration $T_{\rm obs}$, leading to upper bounds on $\mu$.
Finally, we conclude in section \ref{sec:6}.

\section{Cosmic string bursts}
\label{sec:2}

\begin{figure}
    \includegraphics[width=.4\textwidth]{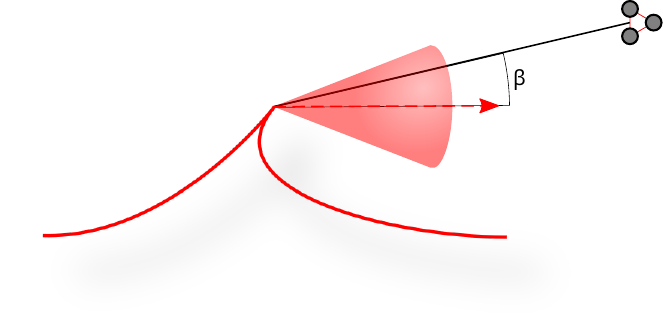}
    \caption{Schematic view of a cosmic string burst, with the beaming angle $\beam$ in red and the misalignement angle $\beta$.}
    \label{fig:h-cusp}
\end{figure}

We start with a brief description of the GWs emitted by cosmic string cusps,
namely points on the string which travel instantly at velocities close to the speed of light, see~\cite{Damour:2000wa,Damour:2001bk,Binetruy:2009vt} for detailed calculations.

The emission from these strong GW sources is concentrated in a beam, see \cref{fig:h-cusp}, with a half-angle
\begin{equation}
    \label{eq:cusp beaming angle}
    \beam(f) = \left[g_2 f (1+z) \ell\right]^{-1/3},
\end{equation}
where $\ell$ is the invariant length of the loop at redshift $z$ containing the cusp, $f$ is the observed GW frequency, and $g_2$ is a $\order{1}$ coefficient that we fixed to $\sqrt{3} / 4$ as derived in \cite{Damour:2001bk,Siemens:2006vk}.
Note that the beaming angle is limited to $\beam(f) < 1$.
The Fourier transform of the cusp waveform is spread over a wide range of frequencies following a power-law $\tilde{h}(f) \sim A f^{-4/3}$.
Its amplitude is given by
\begin{equation}
    A(\ell,z,\mu) = g_1 \frac{G\mu \ell^{2/3}}{(1+z)^{1/3}r(z)},
    \label{eq:Al}
\end{equation}
where $r(z)$ the proper distance to the cusp, and $g_1 \approx 0.85$.
In fact, the signal is cutoff at low frequencies by the fundamental frequency of the loop $f_0 = 2 / \ell$, which in the detector frame imposes
\begin{equation}
    f > \flow \equiv \frac{2}{\ell (1 + z)}.
\end{equation}
Since the beaming angle $\beam$ becomes narrower as the frequency increases, see \cref{eq:cusp beaming angle}, any misalignment of the observer by a small angle $\beta$ from the cusp direction results in a cutoff at high frequencies when $\beta > \beam$.
Hence the observed frequency must satisfy
\begin{equation}
    f < \fhigh \equiv \frac{1}{g_2 \ell \beta^3 (1 + z)}.
\end{equation}

\begin{figure}
\label{fig:realh}
\includegraphics[width=.45\textwidth]{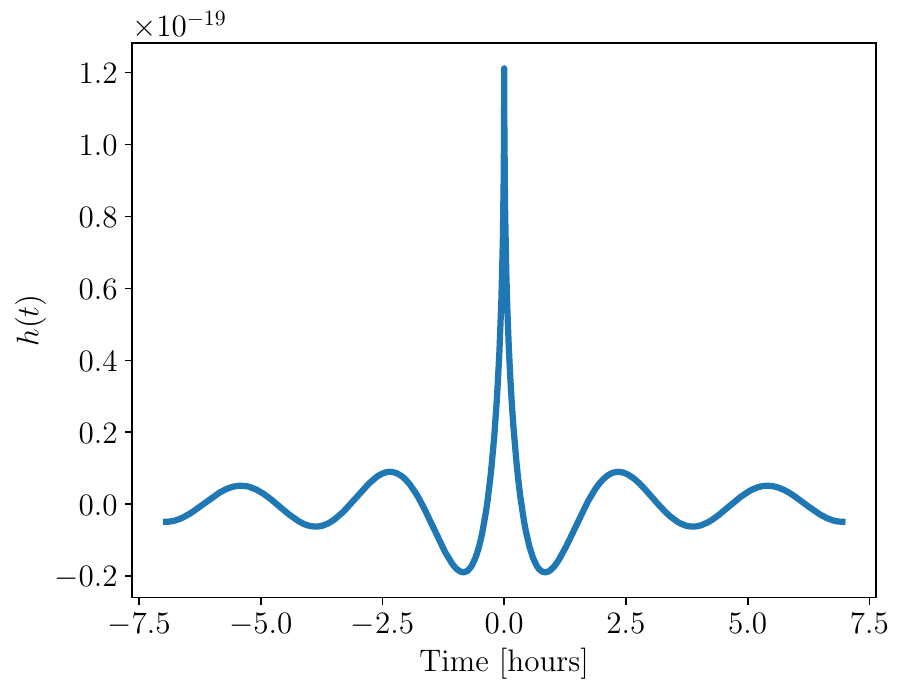}
    \caption{Cusp strain in time domain computed using \cref{eq:cusp-time}, and fixing (see Section \ref{sec:snr}) $\flow=f_1 = 0.1$mHz, $\fhigh=f_2=50$mHz, characteristic of LISA.
    }
\end{figure}

As a consequence, and as the GW signal is linearly polarized, the waveform of a cusp is only characterized by
\begin{equation}
    \label{eq:cusp-frequency}
    \tilde{h}(f) = A |f|^{-4/3} \Theta(f - \flow) \Theta(\fhigh - f),
\end{equation}
which can also be expressed in the time domain with a real Fourier transform
\begin{equation}
    \label{eq:cusp-time}
    h(t) = 2 A \int_{\flow}^{\fhigh} f^{-4/3} \cos (2\pi f t) \dd{f}.
\end{equation}
This is plotted in Fig.~ \ref{fig:realh} where, for illustrative purposes, we have chosen values of $\flow$ and $\fhigh$ characteristic of the LISA sensitivity band, see Section \ref{sec:snr}.
Finally, we choose the convention that for a polarization angle $\psi$ we have in the solar system barycentre frame,
\begin{equation}
    \label{eq:h_plus_cross}
    h_{+}(t) = \cos(2\psi)h(t) \textup{ and } h_{\times} (t) = \sin(2\psi)h(t).
\end{equation}

\section{LISA response}
\label{sec:snr}

LISA has a non-trivial response to the GW signal.
Not only is the wavelength of the GWs comparable to the armlength, but also time-delay interferometry (TDI) must be used.
LISA's satellites follow geodesic motion around the sun and, as a result, the distance between them is not equal and slowly changes in time (breathing and flexing). TDI removes the laser frequency noise by delaying and recombining individual measurements {along the links connecting the spacecrafts} to reproduce the differential measurement with an equal optical path (see \cite{Tinto:2020fcc} and references therein {for more details}). { Combining the measurements in each pair of arms gives us
3 Michelson TDI datasets referred to as $X$, $Y$ and $Z$. }

{The effective duration of the GW burst from cosmic strings is set by the lowest frequencies the gravitational wave detector can detect.
For LISA, $\flow \sim 10^{-4}$ Hz which leads to an effective duration of $10^4$ seconds. This} is therefore much shorter than
the orbital motion of LISA.
With a very high precision (as we will justify later by working in time domain), we {can thus} consider LISA as static (``frozen"), fixing its position at the maximum of the GW amplitude in the time domain.
{With those assumptions,} the response becomes a function of angular frequency $\omega =2\pi f$ only, and the Michelson $X$-TDI combination is given by (see Eq.~(32) of \cite{Babak:2021mhe})
\begin{equation}
 \tilde X_{S} = \omega L \sin(\omega L) e^{i \omega L} \tilde{h}(\omega/2\pi) \left[ F^+_{13}\Upsilon_{13} - F^+_{12}\Upsilon_{12} \right] ,
 \label{Xstat}
\end{equation}
where the subscript $S$ indicates the static-LISA approximation and the other two Michelson combinations, {$Y$ and $Z$,} can be obtained by the permutation of spacecraft indices $1\to 2\to 3 \to 1$.  Note that in computing the response, we can safely assume equal armlengths,  $L=L_{12}=L_{23}=L_{31}$, the precise armlength measurement is required mainly for the laser frequency cancellation.  The $F$ and $\Upsilon$ functions, see \cite{Babak:2021mhe}, depend on the geometry and position of LISA and the polarization angle $\psi$. Note that this expression corresponds to 1.5-TDI generation \cite{Tinto:2020fcc}.  {Each} Michelson combination shares one link and, therefore,  contains correlated noise. {By a linear combination of $X$, $Y$, $Z$}, one can form noise-orthogonal (uncorrelated) datasets referred to as $A$, $E$, $T$, see for example \cite{Tinto:2020fcc}. Since the response strongly suppresses the presence of a GW signal in the $T$-combination, we compute the signal-to-noise ratio (SNR) using only $A$ and $E$.

Finally, we use the power spectral density {$S_A(f) = S_E(f)$} of the LISA noise given in \cite{Babak:2021mhe}. This includes the contribution of galactic confusion noise, for which we have chosen the nominal time span of the LISA mission $T_{\rm{obs}} = 4.5$ years.
Note that the noise rises sharply at low frequencies (below 0.1mHz) and at high frequencies (above 0.2 Hz).
The SNR is then computed in the usual way
 \begin{equation}
 \textup{SNR}^2 = 4 \Re  \int_{f_1}^{f_2} \frac{{|\tilde{A}(f)|}^2 + {|\tilde{E}(f)|}^2}{S_A(f)}\, \dd{f}.
 \label{eq:snr}
 \end{equation}
 We have chosen $f_{1} = 0.1$mHz and $f_{2} = 50$mHz reflecting the LISA sensitivity band.

As an additional check, we have also generated the signal in the time domain, using \cref{eq:h_plus_cross} and following the procedure described in \cite{Babak:2021mhe}. Namely, we first compute the response to the GW burst for a single laser link: from the sender ($s$) to the receiver ($r$), using
\begin{equation}
    \label{eq: single link response}
    y^{GW}_{rs} = \frac{\Phi_{rs}(t_s - \Vec{k}\cdot\Vec{R_s}(t_s)) - \Phi_{rs}(t-\Vec{k}\cdot\Vec{R_r}(t))}{2(1 - \Vec{k}\cdot\Vec{n}_{rs})},
\end{equation}
where $\Vec{R}_{s/r}$ is the vector position of the sending/receiving spacecraft, $\Vec{n}_{rs}$ is a unit vector along the sender-receiver link,
 $\Vec{k}$ corresponds to the direction of propagation of the GW and $\Phi_{rs}$ is the projection of the GW strain on the link $\Phi_{rs} = \Vec{n}_{rs}^i\Vec{n}_{rs}^jh_{ij}$. We then computed the TDI combinations using their definition (by applying the time delays of Eq.~(14) in \cite{Babak:2021mhe} to \cref{eq: single link response}):
\begin{align}
   &X_{1.5} =
y_{13} + D_{13}  y_{31} + D_{13}  D_{31} y_{12} + D_{13}  D_{31} D_{12} y_{21} \nonumber  \\
-&  \left(y_{12} + D_{12} y_{21} + D_{12} D_{21} y_{13} + D_{12} D_{21} D_{13}  y_{31} \right), \label{eq:Xtdi}
\end{align}
 where we have used the short-hand notation for the delay operator $D_{ij} x(t) = x(t - L_{ij}/c)$. This is the Michelson TDI-1.5 combination without any approximations. {After calculating the Fourier transforms of $A$ and $E$ numerically}, we have evaluated the SNR according to \cref{eq:snr} using the full TDI and have confirmed the validity of the static LISA approximation  \cref{Xstat}.
{From a practical point of view we consider the TDI combinations, which contain the GW signal together with  the instrumental response, as LISA's data. It is given either by \cref{Xstat} in frequency domain or by \cref{eq:Xtdi} in time domain.}

Due to its finite sensitivity, LISA can only detect a fraction of the cosmic string burst directed at the instrument.
We assess the {\it detection efficiency} of LISA using $P(\textup{SNR} > x | A, \beta, z)$, the probability that the SNR of a GW burst with amplitude $A$, misalignment angle $\beta$ at redshift $z$ is higher than a given value $x$. We will calculate it in the following section.

\begin{figure*}
	\includegraphics[width=.45\textwidth]{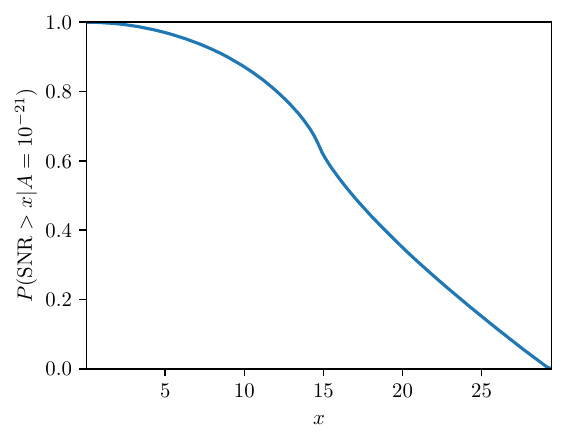}
	\includegraphics[width=.45\textwidth]{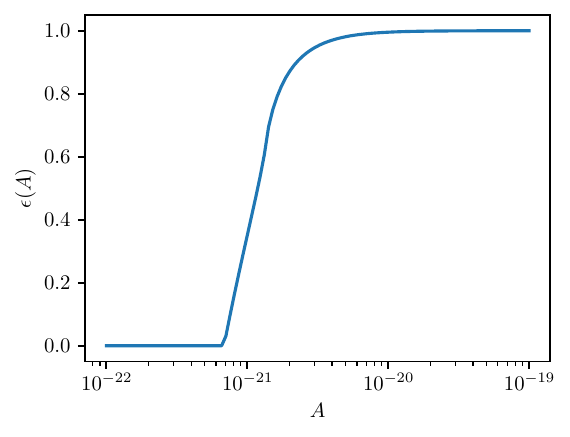}
    \caption{
        Left panel: Detection efficiency of LISA for a burst of amplitude $A=10^{-21} \mathrm{s}^{-1/3}$ marginalized over the sky-localization of the source and polarization angle.
        Right panel: Probability that a burst with amplitude $A$ has SNR larger than $\snrcut = 20$.
    }
    \label{fig:P}
\end{figure*}

\section{\label{sec:rate}Rate of burst in LISA}

Inspired by the framework established in Ref.~\cite{Siemens:2006vk}, we first calculate the event rate $\Psi$ for an idealised `perfect' observer who can detect any signal, however weak.
This rate $\Psi$ is given in terms of the number of bursts $\nu$ that are emitted per cosmic time, per proper volume $V(z)$, per unit angle $\beta$ and per unit loop length $\ell$:
\begin{equation}
    \label{eq:psi}
    \Psi(\ell, \beta, z) = \frac{1}{1+z} \frac{\partial^4 \nu}{\partial\beta \partial t \partial \ell \partial V} = \frac{\sin\beta}{(1+z)\ell} N_c \pdv{\mathcal{N}}{\ell}{V},
\end{equation}
where we have introduced the average number of cusps per loop oscillation $N_c$ and the loop number density $\pdv*{\mathcal{N}}{\ell}{V}$.
In this paper, we consider two models for the loop number density, the BOS \cite{Blanco-Pillado:2013qja, Blanco-Pillado:2017oxo} and LRS \cite{Ringeval:2005kr, Lorenz:2010sm} models.
These models were considered within the LISA collaboration~\cite{Auclair:2019wcv,LISACosmologyWorkingGroup:2022jok} and the LVK collaboration~\cite{LIGOScientific:2017ikf,LIGOScientific:2021nrg}, and the explicit expressions for $\pdv*{\mathcal{N}}{\ell}{V}$ may be found in the references above.
Both models aim at describing the population of sub-Hubble loops in the universe, hence they are only valid in the range
\begin{equation}
    \ell < \alpha t(z),
    \label{eq:size}
\end{equation}
with $\alpha = \order{0.1}$.

In order to make the connection with \cref{sec:snr}, we now express $\Psi$ in terms of amplitude using \cref{eq:Al}, namely
\begin{equation}
    \Psi(A, \beta, z)
    = \abs{\pdv{\ell}{A}} \Psi(\ell, \beta, z)
    =  \frac{3 N_c}{2A}\frac{\sin\beta}{(1+z)} \pdv{\mathcal{N}}{\ell}{V}.
\end{equation}
The fraction of events per unit time detected by LISA is $\Psi$ weighted by the detection efficiency of LISA,
\begin{equation}
    \label{eq:lisa-rate}
    \Rlisa = \int \dd{z} \dd{A} \dd{\beta} \Psi(A, \beta, z) P(\textup{SNR} > \snrcut | A, \beta, z).
\end{equation}
For simplicity, we now assume that the SNR of the burst is entirely determined by its amplitude $A$,
namely
\begin{equation}
    \label{eq: efficiency_approx}
    P(\textup{SNR} > \snrcut | A, \beta, z) \sim P(\textup{SNR} > \snrcut | A).
\end{equation}
This is an exact statement for bursts that are perfect power-laws in the frequency band of LISA
\begin{equation}
    \flow < f_1 < f_2 < \fhigh.
    \label{eq:rule}
\end{equation}
We therefore take the conservative approach to discard all the bursts that do not satisfy \cref{eq:rule}.
Note that the choice of the arbitrary frequencies $f_1$ and $f_2$ has two competing effects on the SNR and the rate of bursts.
Indeed, a wider frequency band would increase the SNR of individual bursts, as can be seen on \cref{eq:snr}.
However, it would also discard a larger number of burst candidates because of the condition in \cref{eq:rule}.
In this analysis, we checked that varying $(f_1, f_2)$ had no strong impact on our results.

The two inequalities in \cref{eq:rule} can be rewritten as
\begin{align}
    1 &< g_2 (1 + z) \ell f_1 \label{eq:rule1} \\
    \beta &< [g_2 \ell (1 + z) f_2]^{-1/3} \equiv \bup(\ell, z). \label{eq:rule2}
\end{align}
where the first, \cref{eq:rule1}, is the requirement that the beam always remain small, $\beam(f) < 1$, for all the frequency that we consider in our frequency band $[f_1, f_2]$.
\cref{eq:rule2} acts as a upper limit $\bup$ for the misalignement angle, and together the inequalities yield
\begin{equation}
    \bup(\ell, z) < \left(\frac{f_1}{f_2}\right)^{1/3} \approx 0.1.
\end{equation}
Note that, in earlier analyses such as in Refs.~\cite{LIGOScientific:2017ikf,LIGOScientific:2021nrg}, no distinction was made between $f_1$ and $f_2$, and both were referred to as $f_*$.
In this case, the misalignement angle is only bounded from above by $\bup(\ell, z) < 1$.

With these conditions, the only remaining dependence on $\beta$ in \cref{eq:lisa-rate} is the term $\sin \beta$ which can easily be integrated to give
\begin{equation}
    \Rlisa = \int \dd{z} \dd{A} \frac{3 \bup^2 N_c}{4(1+z) A} \pdv{\mathcal{N}}{\ell}{V} P(\textup{SNR} > \snrcut | A),
    \label{eq:last-equation}
\end{equation}
using the approximation $1 - \cos \bup \approx \bup^2 / 2$ since $\bup = \order{0.1}$.

We determine the LISA's efficiency  \eqref{eq: efficiency_approx} for a fixed burst amplitude
$A$ as a fraction of sources distributed uniformly on the sky and in polarization angle detectable with the SNR greater than $x$,  $P(\textup{SNR} > x | A)$. The result is shown in the left panel of \cref{fig:P} for the amplitude $A=10^{-21}$. On the other hand, we can compute the efficiency as a function of the burst amplitude while choosing the SNR threshold $\snrcut =20$. The results are shown in
the right panel; we start detecting the bursts starting with $A \ge 8\times 10^{-22}$.
 The SNR threshold $\snrcut =20$ was chosen based on the simple background estimation. We have  performed a matched filtering on the simulated LISA data containing Galactic white dwarf binaries and instrumental noise (but no bursts from cosmic strings). We have found no events above SNR 17, justifying the choice of our threshold. However, a more exhaustive study using a broad prior on the bursts parameter and realistic simulated data (with other GW sources) is required to establish the  definitive value of $\snrcut$.
Finally, we integrate \cref{eq:last-equation} numerically, enforcing the conditions \cref{eq:size,eq:rule1} in order to obtain $\Rlisa$.

\section{Results}
\label{sec:5}

\begin{figure}[t]
    \includegraphics[width=.475\textwidth]{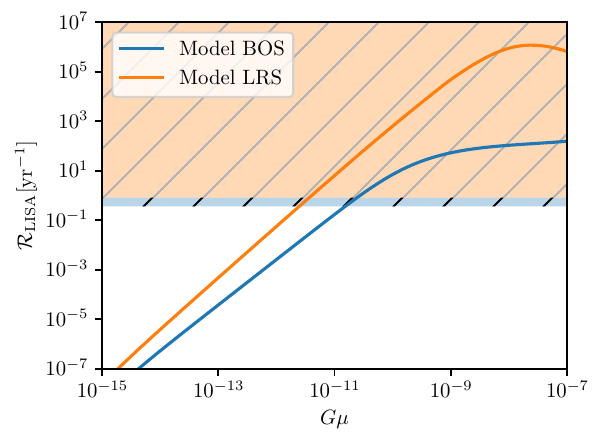}
    \caption{
    Expected rate of detected bursts in LISA as a function of the string tension for models BOS and LRS.
    In case LISA does not detect bursts from cosmic string cusps, the orange hatched region is excluded after $\Tobs = 82 \% \times 4.5$ years and the blue hatched region is excluded after $\Tobs=82 \% \times 10$ years.}
    \label{fig:rate}
\end{figure}

The expected rate of detected bursts in LISA for the BOS and LRS models are presented in \cref{fig:rate} for the fixed number of cusps per oscillation period\footnote{For a loop of length $\ell$, this corresponds to a rate of GW emission $\dot{\ell}=-\Gamma G\mu$ with $\Gamma \approx 50$ \cite{Ringeval:2017eww,Auclair:2019wcv}.} $N_c = 2$.
We compute the expected detection rate for the fixed value of string tension: $G\mu = 10^{-10.1}$ for BOS model and $10^{-10.6}$ for LRS. This tension is compatible with the latest PTA results if we assume that the observed common red noise signal is a stochastic GW signal originating from the string network \cite{EPTA:2023hof}.
The rate for the two models is
\begin{align}
    \Rlisa\qty(G\mu = 10^{-10.1}) &\underset{\mathrm{BOS}}{=} 4 \,\mathrm{yr}^{-1}\\
    \Rlisa\qty(G\mu = 10^{-10.6}) &\underset{\mathrm{LRS}}{=} 30 \,\mathrm{yr}^{-1}.
\end{align}

In the case in which LISA does {\it not} detect bursts from cosmic string cusps during the mission duration $\Tobs$, one can put upper bounds on the string tension.
If we assume that the probability $P(n, \Tobs, G\mu)$ to observe $n$ bursts during $\Tobs$ follows a Poissonian rate with mean $\Tobs \Rlisa(G\mu)$, \emph{i.e.}
\begin{equation}
    P(n, \Tobs, G\mu) = \frac{[\Tobs \Rlisa(G\mu)]^n}{n!} e^{-\Tobs \Rlisa(G\mu)},
\end{equation}
we exclude values of the string tension for which the probability of non-detection ($n=0$) is smaller than $5\%$
\begin{equation}
    \Tobs \Rlisa(G\mu) > -\ln(5\%) \approx 2.99573.
    \label{eq:poisson-5percent}
\end{equation}
Given the shape of the rate $\Rlisa(G\mu)$, see \cref{fig:rate}, the constraint of \cref{eq:poisson-5percent} provides an upper bounds on the string tension.
It is also clear that the bounds on the string tension $G\mu$ will depend on the mission's operating time. The shaded area in \cref{fig:rate} intersecting the expected rate indicates the upper bound on the tension.

We consider two LISA observation scenarios each with a $82\%$ duty cycle: (i) Nominal mission duration  of $4.5$ years, and (ii) Extended mission duration of $10$ years.
In the case of no detection, we will be able to set the  constraints on $G\mu$ for nominal and extended mission periods as given in \autoref{tab:nondetc}.\\
\begin{table}[H]
    \begin{ruledtabular}
    \begin{tabular}{ccc}
         & \textbf{Nominal} & \textbf{Extended} \\
        BOS Model & $G\mu < 3\times10^{-11}$ & $G\mu < 2\times10^{-11}$\\
        LRS Model & $G\mu < 4\times10^{-12}$ & $G\mu < 3\times10^{-12}$
    \end{tabular}
    \end{ruledtabular}
    \caption{$95\%$ confidence upper bound on the string tension from the non-detection of GW cosmic string cusp event for 'Nominal' (4.5 years) and 'Extended' (10 years) mission duration and duty cycle 82\%.}
    \label{tab:nondetc}
\end{table}

\section{Discussion}
\label{sec:6}

We have assessed the capability of the most recent configuration of LISA to detect GW bursts originating from the cosmic string cusps.
We have confirmed the validity of the "frozen" LISA approximation (\cref{Xstat}) by comparing the results with the full LISA response calculations.

As such, this work completes previous analysis of GW signals from cosmic strings that focused mainly on the stochastic GW background or on bursts in the LIGO frequency band.
Whereas the stochastic GW background from strings will be detectable with LISA for $G\mu \gtrsim 10^{-17}$ \cite{Auclair:2019wcv}, we have shown that the GW bursts from the strings with tension $G\mu \gtrsim 10^{-11}$-$10^{-12}$ could be detected with SNR above 20.
{The detection of individual bursts from cosmic strings opens up the opportunity of obtaining the sky-localization of the emitting cosmic string loop~\cite{ShapiroKey:2008ckh} and of complementing other detection methods, such as gravitational microlensing~\cite{Kuijken:2007ma,Bloomfield:2013jka} or electromagnetic counterparts~\cite{Jones-Smith:2009adx,Steer:2010jk}.}

However, we should say that this is not a fair comparison. In order to detect the stochastic GW signal, we need to detect and accurately characterize (to minimize the residuals) all resolvable signals. This is quite a challenging task. On other hand, we need to confirm by a more detailed study the SNR threshold for a reliable detection of astrophysical GW bursts. This threshold will also depend on our abilities to disentangle GW bursts from the instrumental and environmental glitches (noise artifacts).  Some preliminary study was already done in this direction \cite{PhysRevD.99.024019, bayle:tel-03120731} which use the different way glitches and GW signals impact the TDI.

The current bounds on the string tension set by the several PTA collaborations are $G\mu \lesssim 10^{-10}$, which is higher than what is required for detectable bursts, therefore leaving  a window  for the discovery of strings in the LISA band.
In the next decade that remains before the launch of LISA, bounds on $G\mu$ from PTA experiments are likely to become more stringent or to raise great excitement if the common-red-process is confirmed to be a stochastic background of GWs.

\section*{Acknowledgments}
We thank Chiara Caprini, Gijs Nelemans and Antoine Petiteau for their helpful comments and suggestions.
S.B. and H.Q-L acknowledge support from ANR-21-CE31-0026, project MBH\_waves and support from the CNES for the exploration of LISA science
H.Q-L thanks Institut Polytechnique de Paris for funding his PhD.
The work of P.A. is supported by the Wallonia-Brussels Federation Grant ARC \textnumero~19/24-103.

\bibliography{biblio}

\end{document}